
\input harvmac
\def\frac#1#2{{#1\over#2}}

\def\half{\frac12}
\def\ceff{c_{\rm eff}}
\def\zst{Z_{\rm string}}
\def\pprime{p^\prime}

\font\cmss=cmss10 \font\cmsss=cmss10 at 7pt
\def\IZ{\relax\ifmmode\mathchoice
{\hbox{\cmss Z\kern-.4em Z}}{\hbox{\cmss Z\kern-.4em Z}}
{\lower.9pt\hbox{\cmsss Z\kern-.4em Z}}
{\lower1.2pt\hbox{\cmsss Z\kern-.4em Z}}\else{\cmss Z\kern-.4em Z}\fi}

\Title{\vbox{\hbox{PUPT--1334}\hbox{\rm hep-th/9207064}}}
{{\vbox {\centerline{Irreversibility of the Renormalization Group Flow}
\medskip
\centerline{in Two Dimensional Quantum Gravity}
}}}

\centerline{\it David Kutasov}
\smallskip\centerline
{Joseph Henry Laboratories}
\centerline{Princeton University}
\centerline{Princeton, NJ 08544, USA}
\vskip .2in

\noindent
We argue that the torus partition sum in $2d$ (super) gravity, which counts
physical states in the theory, is a decreasing function of the renormalization
group scale. As an application we chart the space of $(\hat c\leq1)$ $c\leq1$
models coupled to (super) gravity, confirming and extending ideas due to
A. Zamolodchikov, and discuss briefly string theory, where our results imply
that the number of degrees of freedom decreases with time.

\Date{7/92}
%
{\bf 1.} The $c$-theorem due to A. Zamolodchikov
\ref\cthem{A. B. Zamolodchikov JETP Lett. {\bf 43} (1986) 730;
see also J. Cardy
and A. Ludwig, Nucl. Phys. {\bf B285[FS19]} (1987) 687;
A. Cappelli, D. Friedan and J. Latorre, Nucl. Phys. {\bf B352} (1991) 616.}
has had a profound influence on our understanding of two dimensional
Quantum Field Theory ($2d$ QFT). In its original formulation it states
that there exists a function $c(t)$
given by a certain combination of two point correlators of stress tensors
$T_{\mu\nu}$, which has the property (in a unitary $2d$ QFT) that it is
a monotonically decreasing function of the Renormalization
Group scale $e^t$. At the fixed
points of the RG (where ${\partial c\over\partial t}=0$)
$c(t)=c$ coincides with the central charge of the appropriate Virasoro
algebra
\ref\bpz{A. Belavin, A. Polyakov and A. B. Zamolodchikov, Nucl. Phys.
{\bf B241} (1984) 333.}. An obvious consequence is that $c(UV)\equiv
c(t\rightarrow-\infty)$ and $c(IR)\equiv c(t\rightarrow\infty)$
always satisfy (in unitary QFT) the inequality:
\eqn\cth{c(UV)\geq c(IR)}
with equality holding if and only if all $\beta$ functions vanish along the
trajectory.

The interpretation given to the $c$-theorem and in particular
to eqn. \cth\ by Zamolodchikov (see also
\ref\kas{D. Kastor, E. Martinec and S. Shenker, Nucl. Phys. {\bf B316}
(1989) 590.}
\ref\dn{D. Kutasov and N. Seiberg, Nucl. Phys. {\bf B358}
(1991) 600.})
has been in terms of {\it thinning of degrees of freedom}
(d.o.f.) as we flow from
the UV to the IR. Indeed, the central charge $c$ of a conformal unitary
$2d$ QFT is related to its number of d.o.f.: consider the canonical
partition sum,
\eqn\zb{Z(\beta)={\rm Tr}\; e^{-\beta H}}
Taking space to be a circle of circumference $2\pi$ (using conformal
invariance), the spectrum of $H$ is related to the conformal dimensions
by
$E=2\pi(\Delta+\bar\Delta-{c\over12})$.
Formally, as $\beta\rightarrow0$ each state is counted once,
so $Z(\beta\rightarrow0)$ gives the number of states. Actually one has to
be slightly more careful: the conformal QFT has an infinite number of states,
hence $Z(\beta\rightarrow0)$ diverges. The rate of this divergence,
which can be taken as a measure of the number of d.o.f. of the system,
is related by modular invariance ($Z(\beta)=Z({1/\beta})$) to the behavior
at $\beta\rightarrow\infty$, where $Z$ is dominated by
the lowest lying eigenstates of $H$.
Since in a unitary CFT the lowest lying state is the identity with
$\Delta=\bar\Delta=0$ and energy $H=-2\pi{c\over12}$, one finds:
\eqn\smallb{Z(\beta\rightarrow0)\sim A\beta^r\exp({\pi c\over6\beta})}
We see that $c$ determines the exponential growth of the number of states
in the CQFT and \cth\ is the statement that the number of states at the
UV fixed point is (exponentially) larger than that at the IR fixed point.

Notice that we have only related $c(t)$ to the density of states at the
(conformal) fixed points  at which $\partial_t c=0$. It is still unclear
whether and in what sense Zamolodchikov's $c(t)$ measures the number
of d.o.f. along the RG trajectory (i.e. in massive theories).

Unitarity was used quite weakly in the foregoing discussion; the main
point of \cth\ was the intuition of the
thinning of degrees of freedom as the system evolved to larger
and larger scales, and unitarity
was only used to determine
the energy of the lowest lying eigenstate of $H$,
which was
related by a modular
transformation to
the density of states.
This observation prompted many people
(see e.g. \dn\
and references therein) to conjecture that there exists
a generalization of the $c$-theorem \cth\ which is valid
in the non-unitary case as well.
The central charge $c$ has to be replaced by $\ceff$ defined
as in \smallb\ by studying $Z(\beta\rightarrow0)\simeq \exp(
{\pi\ceff\over6\beta})$. By modular invariance it is easy to show that:
\eqn\ceffec{
\ceff  = -{6 E_{\rm min} \over \pi} = c-24 \Delta_m;\;\;\;\;
\Delta_m ={1\over 2} {\rm min} (\Delta +\bar \Delta)
}
It is plausible that \cth\ should be generalized to a ``$\ceff$ theorem":
\eqn\ceffth{\ceff(UV)\geq\ceff(IR)}
What seems crucial for the state counting argument leading to
\ceffth\ is that the trace \zb\ be an ordinary trace with all states
contributing with positive signs to $Z(\beta)$. This is very natural, but
there are examples in $2d$ statistical physics where this is not
the case (related to polymers). This restriction will also be significant
when we discuss the case of QFT on fluctuating surfaces where analogous
states contributing with minus signs to \zb\ are sometimes very important.

An example of the $\ceff$-theorem \ceffth\ is the minimal models of \bpz,
where $c=1-{6(p-\pprime)^2\over p\pprime }$, $\ceff=1-{6\over p\pprime }$.
The RG flows lower $p$ and/or $\pprime$ and hence obey \ceffth.
Finally, before turning to gravity, we note that the ``$\ceff$-theorem''
has not been proven so far. In fact, as in the unitary case, it is non
trivial to define the density of states for the massive theories.

\medskip

{\bf 2.} Following Zamolodchikov's work many people have studied
the relevance and implications of the $c$-theorem for string theory
and $2d$ quantum gravity
\ref\gfs{See e.g. A. Polyakov, {\it Gauge Fields and Strings}, Harwood
Academic Publishers (1987).}
but no coherent picture has been developed.
The purpose of this note is to propose
a generalization of \ceffth\ to $2d$ (super) gravity. We will present
a qualitative discussion supported by a few non trivial checks, leaving
proofs and details to a separate publication.

The general situation we envisage is the following. We are handed
a ``matter'' theory with action $S_m(\lambda_i, g_{ab})$ where $\{\lambda_i\}$
is a set of coupling constants defining the theory and $g_{ab}$
the metric of the Riemann surface on which the matter lives. For fixed
$g_{ab}$ the discussion in the previous section applies. Now consider
integrating over the metric $g_{ab}$. This gives rise to a system
of matter coupled to quantum gravity, which is traditionally \gfs\ treated
in conformal gauge $g_{ab}=e^\phi\hat g_{ab}$. The conformal
factor $\phi$ is in general a fluctuating, quantum field; its dynamics
is largely determined by the requirement that the matter plus $\phi$
system be a CFT with $c=26$; for values of the couplings $\lambda_i$
for which the matter system is conformal with central charge $c=c_m$,
$\phi$ is described by the Liouville action \gfs\ with $c=26-c_m$;
in general the dynamics of $\phi$ and matter is more entangled; it
is sometimes useful to treat it by perturbing away from a conformal point.

We would like next to generalize the discussion of the previous section
to the case of fluctuating metric $g_{ab}$,
and identify an object which, like $\ceff$
there, counts physical states. The first point we need to address
is the definition of the RG scale; since all scales are defined in terms
of the dynamical metric $g_{ab}$, which is fluctuating, it seems at first
that the parameter $t$ we had before does not make sense any more. In fact,
there is a scale in the problem. As is well known
\ref\zz{A. B. Zamolodchikov, Phys. Lett. {\bf 117B} (1982) 87.}
the dynamics of the metric poses a well defined problem only when we set
a scale in the path integral. This can be done in a number of ways; the one
most useful to us here is fixing the ``area'': i.e. choose a relevant
microscopic matter operator $V_m$, dress it by the appropriate
Liouville ($\phi$) dependence, generically giving $V_m\exp(\beta_m\phi)$
and insert in the path integral $\delta(A-\int d^2\sigma
V_m\exp(\beta_m\phi))$.
Two useful choices for $V_m$ are $V_m=1$ (in which case $A$ is indeed
the physical area), and $V_m=$ lowest dimension operator in a matter
CFT (which can then be perturbed to the appropriate point in coupling
space $\{\lambda_i\}$, as mentioned above). All correlation functions
in the quantum gravity theory will depend on $A$. Clearly,
small area ($A\rightarrow0$)
is the UV limit, while large area ($A\rightarrow\infty$) is the IR one.
RG flow is the
evolution from small to large area.

Now that we have a scale, we need a definition of the equivalent of
the canonical partition sum \zb. We could write \zb\ with the trace running
over the reparametrization ghosts, and the Liouville field plus matter
$c=26$ CFT, but that would miss the fact that only states with $L_0=
\bar L_0$ are physical in $2d$ quantum gravity. To incorporate
this, we define
\eqn\gb{g(\beta)=\int_{-\half}^\half d\tau_1 {\rm Tr}\; e^{-\beta H+2\pi
i\tau_1(L_0-\bar L_0)}={\rm Tr}\; e^{-\beta H}\delta (L_0-\bar L_0)}
with the trace running over matter, $\phi$ and ghost degrees of freedom.
The quantity $g(\beta\rightarrow0)$ is a measure of the number of degrees of
freedom of the theory and we expect it to decrease under RG flows\foot{
As in the discussion of $\ceff$ on fixed background, we restrict
here to the case where all states contribute with positive sign to \gb.
We will return to the more general case later.}. Now what is the behavior
of $g(\beta)$ as $\beta\rightarrow0$? Generically
$g(\beta)$ diverges exponentially as $\exp({a\over\beta})$. It was shown
in \dn\ that when $a>0$ the quantum gravity system is unstable;
in string theory language, the theory contains tachyons which destabilize
the vacuum. In cases where tachyons are not present we have \dn\
(up to uninteresting constants):
\eqn\om{\lim_{\beta\rightarrow0}g(\beta)=\Omega}
where $\Omega$ is the genus one path integral given by
\eqn\defom{\Omega=A\int[{\cal D}g_{ab}][{\cal D}({\rm matter})]
\exp\left[-S_m(\lambda_i, g_{ab})\right]\delta(A-\int V_m)
=f(\lambda,A)\int_{\cal F}{d^2\tau\over\tau_2^2}\zst(\tau; A)}
where
$$\zst(\tau_1,\tau_2;A)={\rm Tr}\;e^{-2\pi\tau_2(L_0+\bar L_0)+2\pi i\tau_1
(L_0-\bar L_0)}$$
$f(\lambda,A)$ is a certain function of the couplings and scale, and
$\cal F$ is the fundamental domain of the modular group ${\cal F}=\{
\tau=\tau_1+i\tau_2| \tau_2>0,\;
|\tau_1|\leq\half,\;
|\tau|\geq1
\}$.

To recapitulate, the torus path integral \defom\ in $2d$ quantum gravity
measures the number of d.o.f. of the system. It depends on the couplings
$\{\lambda_i\}$ in the matter Lagrangian and on the scale $A$, $\Omega=
\Omega(\lambda, A)$. Hence it is defined along the RG trajectory
$0\leq A<\infty$. It is natural to conjecture, following the logic of the
$c$-theorem, that {\it $\Omega$ decreases along RG trajectories}:
\eqn\cgrav{{\partial\Omega\over\partial\log A}\leq0}
with equality only at fixed points (i.e. when the matter is conformal).
Actually there are subtleties which need to be addressed in order to make
\cgrav\ precise. One such subtlety is the renormalization of the vacuum
energy which is known to occur as $A$ increases; we may have to tune
the effective cosmological constant to zero as a function of the
$\{\lambda_i\}$; this is in any case an ambiguity present in \defom.
Other subtleties have to do with the function $f$ in \defom. We will return
to these and other matters elsewhere; here we will study the consequences
of the analog of \cth, \ceffth\ which is obtained by defining
$\Omega(UV)\equiv\Omega(\lambda, A\rightarrow0)$;
$\Omega(IR)\equiv\Omega(\lambda, A\rightarrow\infty)$, and deducing from
\cgrav:
\eqn\omth{\Omega(UV)\geq \Omega(IR)}
Eq. \omth\ is the main statement of this note. We will not prove it here, but
rather verify its validity in some examples.
Of course, \omth\ is weaker than \cgrav\ but still it will allow us to
discuss quantitatively various RG flow patterns in (super) gravity.

\medskip

{\bf 3.} The discussion so far has been quite abstract.
We now turn to a study of a few test cases. As a first example, consider
a free scalar field $x$ minimally coupled to the metric $g_{ab}$:
\eqn\s{S_m=\int d^2\sigma\sqrt{g} g^{ab}\partial_a x\partial_b x}
We will take $x$ to be compact, $x\sim x+2\pi R$ for later convenience;
the generalization to non compact $x$ is straightforward. As is well known,
after gauge fixing to conformal gauge
$g_{ab}=\hat g_{ab} e^\phi$, the action becomes:
\eqn\sgauge{S_g=\int d^2\sigma\sqrt{\hat g}[(\partial x)^2+(\partial\phi)^2
+2\sqrt{2}\hat R\phi+{\cal L}_{\rm ghost}]}
${\cal L}_{\rm ghost}$ is the ghost Lagrangian whose precise form
is not important for our purposes. The path integral \defom\ is given
by
\eqn\omcone{\Omega=A\int[{\cal D}\phi][{\cal D}x][{\cal D}{\rm ghosts}]
e^{-S_g}
\delta(A-\int \phi e^{-\sqrt{2}\phi})}
On the torus $\hat R=0$; hence by KPZ scaling ${\partial\Omega\over\partial
A}=0$. This path integral has been computed in
\ref\mat{D. Gross and I. Klebanov, Nucl. Phys. {\bf B344} (1990) 639.}
\ref\bk{M. Bershadsky and I. Klebanov, Phys. Rev. Lett. {\bf 65} (1990)
3088; Nucl. Phys. {\bf B360} (1991) 559.}
\ref\st{N. Sakai and Y. Tanii, Int. J. Mod. Phys. {\bf A6} (1991) 2742.}.
The result is (in units $\alpha^\prime=1$):
\eqn\cone{\Omega={1\over24}(R+{1\over R})}
$R$ and $1/R$ are the contributions of the momentum and winding modes
respectively \dn\ \bk.
Of course, since the matter theory is conformal, \cgrav\ and \omth\ are
trivial.
To make the situation more interesting, consider perturbing \s\ to:
\eqn\spert{S_m(\lambda, \rho)=\int d^2\sigma\sqrt{g} \left[
g^{ab}\partial_a x\partial_b x+\sum_{n=0}^N(\lambda_n\cos{n\over R}x+
\rho_n\sin{n\over R} x)\right]}
The conformal dimension of the operator $e^{ipx}$ is $\Delta_p={p^2\over4}$,
so all operators with $n<2R$ ($p<2$) in \spert\ are relevant
(in the RG sense). For a given
radius $R$ there is a finite number of those; defining $N$ to be the
integer which satisfies\foot{We assume $R$ is generic, $R\not\in\IZ/2$,
so the more subtle case of $N/R=2$ does not arise.}:
\eqn\nmax{2R-1<N<2R}
the space of relevant perturbations of the $c=1$ theory \spert\ is
$2N+1$ dimensional.
For any choice of $\lambda, \rho$, the UV fixed point of \spert\ is the
original $c=1$ model since, as usual, in the UV the relevant couplings
$\lambda, \rho\rightarrow0$;
indeed, by KPZ scaling, calculating \omcone\ at area $A$ and couplings
$\lambda_i, \rho_j$ is the same as calculating it at $A=1$ and couplings
$\lambda_iA^{1-{p_i\over2}}, \;\rho_jA^{1-{p_j\over2}}$.
At large distances, $A\rightarrow\infty$, the effective couplings
grow and we flow to a new fixed point which depends on $\lambda_i, \rho_j$.
What is that fixed point?

Consider first the important special case of a single cosine potential
in \spert:
\eqn\vx{S_m=\int d^2\sigma\left[(\partial x)^2+\lambda_n\cos{n\over R}x\right];
\;\;1\leq n\leq N}
Here the potential has $n$ quadratic minima; as $A\rightarrow\infty$
we expect the $x$ field to relax to one of these minima, which
decouple from each other in the infrared limit (mass goes to infinity).
Hence we expect the IR fixed point to describe\foot{The fact that the IR
limit of \vx\ is pure gravity can be verified directly by perturbing the
$c=1$ model in $\lambda$
\ref\greg{G. Moore, Yale preprint YCTP-P1-92 (1992); E. Hsu and D. Kutasov,
unpublished.}.} {\it $n$ decoupled models of pure gravity}.
The partition sum $\Omega$ of pure gravity was calculated
in \bk\ and is equal to $\Omega_{c=0}=1/48$. Thus, for the case of
\vx\ we have:
\eqn\uvir{\Omega(UV)={1\over24}(R+{1\over R});\;\;\;\Omega(IR)={n\over 48}}
But for the cosine \vx\ to be relevant we had to require $n\leq N$ and
\nmax\ $N<2R$. Therefore, $\Omega(IR)={n\over 48}<{2R\over 48}={R\over 24}$,
and in particular,
\eqn\comp{\Omega(UV)=\Omega_{c=1}>\Omega(IR)=n\Omega_{c=0}}
in agreement with \omth.

Eq. \comp\ is an encouraging first sign, but we still have to
analyze the general case \spert\ of an arbitrary potential,
\eqn\potential{V(x)=\sum_{i=0}^N(\lambda_n\cos{n\over R}x+\rho_n\sin{n
\over R}x)}
An important lesson from the special case \vx\ is that in the IR the minima
of the potential \potential\ are dominant; since the barriers separating
the different minima go to infinite height, the system splits up
into a direct sum of non interacting theories at the minima of $V(x)$.
These IR subtheories need not be massive, $c=0$ ones; in general,
by tuning the couplings in \potential\ appropriately one can get
multicritical behavior $V(x)-V(x_0)=C(x-x_0)^{2l}+\cdots$
with $l>1$. Zamolodchikov
\ref\lg{A. B. Zamolodchikov, Sov. J. Nucl. Phys. {\bf 44} (1986) 529.}
has identified the multicritical Lagrangian
\eqn\multi{{\cal L}=(\partial x)^2+x^{2(l-1)}}
(on a fixed background $g_{ab}$) as describing the $(p,\pprime)$ conformal
unitary minimal model \bpz\ with $p=l$, $\pprime=l+1$
(the $(2,3)$ model is massive, $c=0$; $(3,4)$ is Ising; $(4,5)$ tricritical
Ising, etc). This result has never been rigorously proven
(see \kas\ for a discussion), but we will assume it is correct,
and will show below that our results provide further evidence for this
identification.

Returning to \spert, we see that in general the IR fixed point will consist
of a direct sum of different unitary minimal models living at the different
minima of \potential; generically one will find a collection of
pure gravity $(2,3)$ models corresponding to quadratic minima, as in the
discussion of \vx\ above, but by fine tuning one can get different
$(l,l+1)$ models with $l>2$ at different minima.
The partition sum $Z(\lambda, \rho, A)$ of this $2d$ gravity model will
take in the IR the form (to all orders in the genus expansion):
\eqn\interactio{Z(A\rightarrow\infty)\simeq e^{\mu_{\rm eff}A}\left[
\sum_{\rm minima} Z_{(l_i, l_i+1)}(A)+\cdots\right]}
where the $\cdots$ represent interactions between the different minimal
models, which vanish faster than $Z_{(l,l+1)}$ in the IR (presumably
exponentially).
Our task is to show that \omth\ is always satisfied.

To discuss the critical points of \potential\ one can proceed as follows:
define formally the variable $r=\exp({i\over R}x)$. In terms of $r$,
the potential $V$ is:
\eqn\potr{V(r)=\sum_{p=0}^Na_pr^p+{\rm c.c.}}
where $a_p$ are complex coefficients easily related to $\lambda_i, \rho_j$.
Stationary points of $V$, $\partial_x V=0$ are obtained by applying
$\partial_x={i\over R}(r\partial_r-r^*\partial_{r^*})$ to \potr.
We find that the critical points satisfy
\eqn\crit{\sum_{p=0}^N(b_pr^p+b_p^*(r^*)^p)=0}
where $b_p$ are coefficients easily related to $a_p$, whose form will not play
a role below. Since we're  interested in real $x$, we set $r^*=1/r$. Plugging
this in \crit\ and multiplying by $r^N$ we find that the stationary points
of $V$ are given by solutions of a polynomial equation in $r$ of order
$2N$. Hence, in general $V(x)$ \potential\ has $2N$ critical points\foot{
Of course some of those may occur at complex $x$ in which case they are
irrelevant for our purposes (of finding possible IR fixed points).}. E.g.
 in the example of \vx\ (with $n=N$) there are $N$ minima and $N$ maxima
(all quadratic). When one adds to $V(x)=\lambda\cos{N\over R}x$ more general
perturbations \potential, these $N$ minima and maxima move around
as a function of $\{\lambda_i,\rho_j\}$. Multicritical points occur
when two or more stationary points collide at some value of the couplings.
In general, by tuning $n$ coefficients in \potential\ we can reach
the $(n+1, n+2)$ multicritical point \multi\ around a certain minimum,
or a collection of lower ($l\leq n$)
minimal models around several minima of $V$.
The only missing element to check \omth\ in the most general case is the
value of the torus path integral $\Omega_{(n,n+1)}$ for the $(n,n+1)$
unitary minimal model, which is known
\ref\kdv{P. Di Francesco and D. Kutasov, Nucl. Phys. {\bf B342} (1990) 589.}
\bk:
\eqn\omun{\Omega_{(n, n+1)}={n-1\over 48}}
Using the above discussion and \omun\ it is now straightforward to establish
\omth. Start at a point in coupling space where $V(x)$ \potential\
has $N$ quadratic minima and maxima. For such couplings
the result \omth\
is correct \uvir, \comp. Now deform the potential; the positions
of the critical points will shift until some of them collide creating
multicritical points. This can happen in many different ways; e.g.
when two adjacent simple (quadratic) minima enclosing between them a simple
maximum approach each other, they create a new minimum with $V(x)\simeq
x^4$ behavior corresponding \multi\ to an Ising model. The IR partition sum
corresponding to two distinct simple minima is \omun\ $2\Omega_{(2,3)}=2/48$,
the same as that of the Ising model $\Omega_{(3,4)}$. Thus, since the theorem
holds for values of the couplings
for which the IR fixed point includes two pure gravity $(2,3)$ models,
it also holds when they merge to form an
Ising $(3,4)$ model.
More generally, when two adjacent minima of orders $2n$ and $2m$
(with a simple maximum between them) collide, they create a single minimum of
order $2(n+m)$. The IR partition sum {\it does not change} in the process,
since \omun:
\eqn\bosid{\Omega_{(n+1, n+2)}+\Omega_{(m+1, m+2)}=\Omega_{(n+m+1, n+m+2)}=
(n+m)/48}
The number of d.o.f. in the two distinct minima is the same as that in the
single combined multicritical minimum: no d.o.f. are lost, rather they are
``redistributed''.

Another possibility for multicritical behavior arises when a minimum merges
with an adjacent maximum. This creates an inflection point $V(x)\simeq
x^{2n+1}$ which does not store degrees of freedom in the IR (since it is
an unstable critical point); here the number of d.o.f. when the minimum
and maximum are separate is larger than when they coincide, and
again \omth\ holds.
In general, by a sequence of these and other operations, the potential
\potential\ can be made to describe arbitrarily complicated IR behavior
consisting of a collection of many different unitary minimal models (for large
enough $N$ \nmax).
One can convince oneself along similar lines to the above discussion
that  \omth\ always holds. This completes the proof of \omth\ for the
$c\leq1$ system \spert.
The above considerations can be repeated for the case of $c=1$ orbifolds,
with similar conclusions. We will not do that here.

\medskip

{\bf 4.} Our second example is $2d$ supergravity coupled to a scalar
superfield. Some of the qualitative features are similar to the bosonic case
so we'll be brief; still, some new features make it worthwhile to discuss
this model.

The supersymmetric version of \s\ is well known
\ref\ws{S. Deser and B. Zumino, Phys. Lett. {\bf 65B} (1976) 369;
L. Brink, P. Di Vecchia and P. Howe, Phys. Lett. {\bf 65B} (1976) 471.}.
It involves a superfield $X$ and metric and gravitino fields. It is usually
treated in the superconformal gauge \gfs\ where the gravitational sector
is described by the super Liouville mode
$\Phi=\phi+\theta\psi+\bar\theta\bar\psi+\theta\bar\theta F$
(and super reparametrization ghosts). The scale is set by inserting in the
path integral $\delta(l-\int d^2\sigma d^2\theta \Phi e^{-\Phi})$ (compare
to \omcone). $(\sigma, \theta)$ are coordinates on the super Riemann surface.
The parameter $l$ has dimensions of length; the physical area is
$A=l^2$. As before, small $l$ is the UV and RG evolution increases $l$.
The matter action is
\eqn\smsusy{S_m=\int d^2\sigma d^2\theta(DX)^2}
where $D=\partial_\theta+\theta\partial_z$, $X=x+\theta\psi_x+\bar\theta
\bar\psi_x+\theta\bar\theta F_x$, etc. Again we take $x$ to be periodic,
$x\sim x+2\pi R$. There are many distinct ways to couple the superfield
$X$ \smsusy\ to super Liouville, involving different ``GSO projections''
\ref\gsw{For a review see
M.B. Green, J.H. Schwarz and E. Witten, {\it Superstring Theory},
Cambridge University Press (1987).}.
In $2d$ supergravity it is natural to perform a ``non-chiral" GSO projection;
there are different ways to do that (four different theories can be
obtained for generic $R$, and in addition various orbifold theories
and isolated ones can be constructed). We can also perform a chiral
GSO projection
\ref\ks{D. Kutasov and N. Seiberg, Phys. Lett. {\bf 251B} (1990) 67.},
and there are many ways of doing that as well. We will discuss here only
one of the non chiral theories (although the other choices involve new
phenomena). Namely we take $x$, the lower component of $X$, to live
on a circle of radius $R$ as before, and sum over spin structures
of the fermions $\psi_x, \psi$ independently of $x$. This fixes the spectrum
in the NS sector, but as usual there is a $\IZ_2$ ambiguity in the Ramond
sector \gsw. We choose to project out all Ramond winding modes, leaving the
momentum modes intact. The torus path integral for this system is
\ref\dkm{M. Douglas, D. Kutasov and E. Martinec, unpublished.}
given by\foot{This path integral has been also considered in \bk; we find a
different answer for the odd spin structure \dkm. Below we will quote results
for the supersymmetric minimal models, which also differ from \bk\ for
similar reasons.}:
\eqn\susyone{\Omega={1\over12\sqrt{2}}(2R+{1\over R})}
where the $2R$ come (in the sense of \dn, \gb) from the NS and Ramond
momentum modes, and the $1/R$ from the NS winding modes (recall that
there are no Ramond winding modes in this model). As in the
bosonic case, we will also need the results for the unitary supersymmetric
minimal models
\ref\sumi{D. Friedan, Z. Qiu and S. Shenker, Phys. Lett. {\bf 151B} (1985)
37; M. Bershadsky, V. Knizhnik and M. Teitelman, Phys. Lett. {\bf 151B} (1985)
31.}
\kas, which are again labeled by two integers $(p, \pprime)$ with
$|p-\pprime|=2$. Even and odd $p$ models behave differently; for odd $p$
the supersymmetric index of the theory vanishes and supersymmetry can
generically be broken. For even $p$ the index is non zero. The
results for the torus partition sum when these models are coupled to
supergravity are \dkm:
\eqn\omsusy{\eqalign{
\Omega_{(p, \pprime)}=&{(p-1)(\pprime-1)\over12(p+\pprime-2)}
\;\;\;\;\;\;\;\;\;\;\;\;\;\;\;p, \pprime
\;\;{\rm odd}\cr
\Omega_{(p, \pprime)}=&{(p-1)(\pprime-1)+1\over12(p+\pprime-2)}\;\;\;
\;\;\;\;\;\;p, \pprime
\;\;{\rm even}\cr}}
For unitary models
\eqn\unomsusy{
\Omega_{(2n, 2n+2)}={2n\over24};\;\;
\Omega_{(2n-1, 2n+1)}={1\over24}(2n-1-{1\over2n-1})}
Finally, we will utilize Zamolodchikov's results for the supersymmetric
case. He has shown \lg\ that the Lagrangian
\eqn\slg{{\cal L}=(DX)^2+gX^p}
describes the $(p,p+2)$ supersymmetric (unitary) minimal model.

The discussion starts similarly to the bosonic case. Consider
turning on a superpotential $W(X)=\lambda_n\cos{n\over R}X$ in \smsusy.
This defines a relevant perturbation for ${1\over4}({n\over R})^2<\half$
or $n<\sqrt{2} R$. The IR limit is a theory of $n$ $\hat c=0$ $(2,4)$ models.
Hence we have $\Omega(UV)=\Omega_{\hat c=1}={1\over12\sqrt{2}}(2R+{1\over R})$
and $\Omega(IR)=n\Omega_{(2,4)}=n{1\over12}<{\sqrt{2}R\over12}$. As expected,
\omth\ is valid:
\eqn\susomp{\Omega(UV)=\Omega_{\hat c=1}>\Omega(IR)=n\Omega_{\hat c=0}}
At first glance it appears that we have undercounted the d.o.f. of the
infrared theory. Indeed, {\it all} critical points of the superpotential
($W^\prime=0$) correspond in flat space to minimal models. For the cosine
superpotential discussed above, we have ignored the $n$ maxima of $W$,
which also give rise to \slg, with $p=2,\;g<0$
(in flat space the sign of $g$ doesn't matter
since the bosonic potential $(W^\prime)^2$ is stable for both signs).
On the basis of \omth\
we know that these should not contribute to $\Omega(IR)$, but this seems
in contradiction with \slg.
The resolution is that maxima of the superpotential correspond to
unstable critical points in the presence of supergravity.
To see that, consider coupling the matter theory described by
${\cal L}_m=(DX)^2+gX^p$ to the super Liouville mode.
One finds
\eqn\sugra{{\cal L}_{\rm sg}=(DX)^2+(D\Phi)^2+{g\over2}\left[(X+i\Phi)^p+
(X-i\Phi)^p+\cdots\right]e^{-\Phi}}
where the $\cdots$ denote lower powers of $X\pm i\Phi$ whose coefficients
will be fixed later. The form \sugra\ of the interaction Lagrangian
is determined as usual by BRST invariance.
The Lagrangian \sugra\ describes a sigma model with two dimensional target
space $(X^\mu)=(X,\Phi)$ and superpotential
\eqn\wsg{W_{\rm sg}(X^\mu)={g\over2}\left[(X+i\Phi)^p+(X-i\Phi)^p+\cdots\right]
e^{-\Phi}}
To find the potential for the bosonic (lower) components of $X^\mu$,
$x^\mu$, we should evaluate $\int d^2\theta {\cal L}_{\rm sg}$.
In general, one finds,
\eqn\vxmu{\int d^2\theta\left[ (DX^\mu)^2+W_{\rm sg}(X^\mu)\right]=
(\partial x^\mu)^2+\psi^\mu\partial\psi^\mu+(\partial_\mu W_{\rm sg})^2|_{X=x}
+\bar\psi_{\mu}\psi_\nu\partial_\mu\partial_\nu W_{\rm sg}|_{X=x}}
The potential $(\partial_\mu W_{\rm sg})^2$ {\it vanishes} in super Liouville
theory, essentially because of the exponential $e^{-\Phi}$ in the
superpotential\foot{
This is similar (in one higher dimensional target space) to the super
Liouville action itself $\int d^2\theta\left[(D\Phi)^2+\mu \exp(\gamma\Phi)
\right]=(\partial\phi)^2+\psi\partial\psi+\mu\gamma^2
\bar\psi\psi\exp(\gamma\phi)$.
The term $\gamma^2\mu^2 \exp(\gamma\phi)\exp(\gamma\phi)$ \vxmu\ vanishes
due to normal ordering.} \wsg. There is nevertheless a potential, coming
from the last term in \vxmu. At fixed length $l$ one can put $\psi_x=\bar
\psi_x=0$, but not $\psi=\bar \psi=0$. The term $\bar\psi\psi\partial^2_\Phi
W_{\rm sg}|_{\Phi=\phi}$ in \vxmu\ induces a potential for $x$. One can
tune the coefficients of the lower terms in \sugra\ such that the potential
is $V(x)=glx^p$. We see that for even $p$, $x=0$ is an unstable critical
point if $g<0$, while for odd $p$, $x=0$ is always unstable. Therefore
the analysis leading to eq. \susomp\ is justified -- we can ignore the maxima
of $W$.

Now perturb the superpotential $W$ as in \spert,
\eqn\superp{W(X)=\sum_{n=0}^N(\lambda_n\cos{n\over R}X+\rho_n\sin{n\over R}X)}
$N$ is the largest integer which satisfies $N<\sqrt{2}R$. Again
one can follow the motion of the critical points of $W$ as a function of the
couplings $\lambda, \rho$.
The main question is as before what happens when critical
points collide. When adjacent minima get close, no degrees of freedom
are lost as in the bosonic case, since \unomsusy:
\eqn\comb{\Omega_{(2n,2n+2)}+
\Omega_{(2m,2m+2)}=
\Omega_{(2(n+m),2(n+m)+2)}(={n+m\over12})}
Of course, as in the bosonic case \bosid,
eq. \comb\ is to be compared to the behavior
of the superpotential \slg\ as two minima collide: $X^{2n}\cdot X^{2m}=
X^{2n+2m}$.
The only new phenomenon is the behavior near inflection points, since by
\slg\ for odd $p=2n+1$ the critical point carries in the IR a non trivial
theory. We saw above that
after coupling to supergravity, these critical points too
(as the ones corresponding to maxima of $W$) become unstable.
It is interesting that only the even minimal models (those with
non zero supersymmetric index) correspond to stable IR fixed points
after coupling to supergravity. See \kas,
\ref\superkdv{P. Di Francesco, J. Distler and D. Kutasov, Mod. Phys. Lett.
{\bf A5} (1990) 2135.}\ for related speculations.
Since maxima and inflection points
of $W$ are unstable as in the previous section,
the rest of the analysis
is identical to the bosonic case, and as there it confirms \omth.

Before leaving our case studies, we would like to mention one additional
simple check of \omth\ -- the flow between minimal models. It is known that
both in bosonic and in supersymmetric minimal models
there are flows between $(p, \pprime)$ models, which decrease $p, \pprime$.
Indeed $\Omega_{(p, \pprime)}=(p-1)(\pprime-1)/24(p+\pprime-1)$
in the bosonic case \bk\ decreases as $p, \pprime$ decrease. In the
supersymmetric case the models with even and odd $p, \pprime$ flow down
separately \kas\
\ref\superflow{
R. Pogosyan, Sov. J. Nucl. Phys. {\bf 48} (1988) 763;
Y. Kitazawa, N. Ishibashi, A. Kato, K. Kobayashi, Y. Matsuo
and S. Odake, Nucl. Phys. {\bf B306} (1988) 425.},
and it is easy to check that in both cases
$\Omega_{(p, \pprime)}$
\omsusy\ decreases along these flows.

\medskip

{\bf 5.}
One of the reasons for studying $2d$ gravity throughout the years
has been its relevance to string theory \gfs\ \gsw. Models of two dimensional
gravity are vacua of string theory (albeit not the most general ones).
There is a compelling
interpretation of the Liouville coordinate $\phi$ as Euclidean time
\ref\time{J. Polchinski, Nucl. Phys. {\bf B324} (1989) 123;
S. Das, S. Naik and S. Wadia, Nod. Phys. Lett. {\bf A4} (1989) 1033;
T. Banks and J. Lykken, Nucl. Phys. {\bf B331} (1990) 173.}.
The values of the couplings $\{\lambda_i\}$ specifying the matter
theory (e.g. in \spert)
are then interpreted as initial conditions for the various space time fields.
Dressing by Liouville corresponds to propagation in Euclidean time. The fixed
area partition sum \defom, \omcone\ has the interpretation of fixing
the Euclidean time $t\sim\log A$. We are accustomed to fixing values
of coordinates by inserting in the
world sheet path integral $\delta(x_0-\int x)$. Since $\phi$ is not an
ordinary world sheet scalar, this would not be a gauge invariant
procedure for it. Instead, we fix the value of $\phi$ by
a $\delta(e^{\phi_0}-\int e^\phi)$ as in \defom, \omcone.

The main statements \cgrav, \omth\ discussed above assume a space time
meaning in this framework. If Liouville is thought of as time, then we find
that the number of states in string theory decreases with (Euclidean) time.
It is sometimes advantageous to interpret $\phi$ as a space coordinate.
This is the case in two dimensional string theory, some of whose time dependent
solutions we have studied above (e.g. \s, \spert);
an interesting vacuum of that theory is the $2d$ black hole
\ref\black{E. Witten, Phys. Rev. {\bf D44} (1991) 314.}. There, we expect
a number of degrees of freedom which depends on the position in space.
It would be interesting to study the number of states in the $2d$
black hole background as a function of the distance from the horizon. To do
that one can fix the equivalent of the area directly in the coset model
of \black\ and vary it as discussed above. This may provide insight into
the nature of the physics of the horizon and singularity in the model.

Another string theory aspect of our results has to do with string
theories with space time fermions. Two dimensional string theory is after
all only a toy model, because it has very few physical states (by string
standards). In higher dimensional string theory consistency is
achieved by balancing a very large number of fermions (which contribute
with minus sign to \gb) and bosons, such that the effective number of d.o.f.
(bosons minus fermions) is very small \dn. Clearly in such a situation
we do not expect \cgrav, \omth\ to be valid since the number of bosons
and fermions may separately decrease while their difference may well
increase. We can not offer any final results on this issue, but expect
the basic intuition behind \cgrav\ to still be valid in such a situation.
Hence, defining similarly to \gb\ (see also \dn):
\eqn\gbf{
g_B(\beta)={\rm Tr}_B\;e^{-\beta H}\delta(L_0-\bar L_0);\;\;
g_F(\beta)={\rm Tr}_F\;e^{-\beta H}\delta(L_0-\bar L_0)}
where the first trace runs only over bosonic physical excitations and the
second
only over (space time) fermions, we would expect
$\lim_{\beta\rightarrow0}
g_B(\beta)$ and
$\lim_{\beta\rightarrow0}
g_F(\beta)$
to separately decrease along RG trajectories, so that the total
number of states (bosons plus fermions) decreases with RG time $t$.
Of course, $g_B(\beta), \;g_F(\beta)$ separately diverge as $e^{a\over\beta}$
as $\beta\rightarrow0$, and we would have to count states by a procedure
similar to that discussed in equation \smallb.
This will also bring our results
closer to those of Zamolodchikov \cthem.
All this will have to be left for future work.
\bigbreak\bigskip\bigskip\centerline{{\bf Acknowledgements}}\nobreak
I am grateful to M. Bershadsky, E. Martinec and A. Zamolodchikov
for useful discussions.
This work was partially supported
by NSF grant PHY90-21984.

\listrefs
\end